\documentclass[usenatbib]{mn2e}
\usepackage{graphicx,times}
\usepackage{bm}
\newcommand{\EQ}{\begin{equation}}
\newcommand{\EN}{\end{equation}}
\newcommand{\EQA}{\begin{eqnarray}}
\newcommand{\ENA}{\end{eqnarray}}

\newcommand{\EEq}[1]{Equation~(\ref{#1})}
\newcommand{\Eq}[1]{Equation~(\ref{#1})}
\newcommand{\Eqs}[2]{Equations~(\ref{#1}) and~(\ref{#2})}

\newcommand{\App}[1]{Appendix~\ref{#1}}
\newcommand{\Sec}[1]{Section~\ref{#1}}
\newcommand{\Secs}[2]{Sections~\ref{#1} and \ref{#2}}
\newcommand{\Secss}[2]{Sections~\ref{#1}--\ref{#2}}
\newcommand{\Fig}[1]{Figure~\ref{#1}}

\newcommand{\bra}[1]{\langle #1\rangle}

{}
{}
{}

{}
{}
{}
{}
{}
{}
{}
{}
{}
{}
{}
{}
{}
{}
{}
{}
{}

{}
{}
{}

{}

{}
{}

\newcommand{\xxx}{\hat{\mbox{\boldmath $x$}} {}}

\newcommand{\zzz}{\hat{\mbox{\boldmath $z$}} {}}

\newcommand{\RR}{\mbox{\boldmath $R$} {}}
\newcommand{\kk}{\bm{k}}

\newcommand{\xx}{\bm{x}}

\newcommand{\UU}{\mbox{\boldmath $U$} {}}

\newcommand{\BB}{\mbox{\boldmath $B$} {}}
\newcommand{\EE}{\mbox{\boldmath $E$} {}}

\newcommand{\JJ}{\mbox{\boldmath $J$} {}}
\newcommand{\SSS}{\mbox{\boldmath $S$} {}}
\newcommand{\AAA}{\mbox{\boldmath $A$} {}}

\newcommand{\ff}{\mbox{\boldmath $f$} {}}

\newcommand{\FF}{\mbox{\boldmath $F$} {}}

\newcommand{\nab}{\mbox{\boldmath $\nabla$} {}}

\newcommand{\SSSS}{\mbox{\boldmath ${\sf S}$} {}}

\newcommand{\DD}{{\rm D} {}}

\newcommand{\dd}{{\rm d} {}}

\newcommand{\const}{{\rm const}  {}}

\def\Ma{\mbox{\rm Ma}}

\def\Rm{R_{\rm m}}

\def\cs{c_{\rm s}}

\def\kf{k_{\rm f}}

\def\Brms{B_{\rm rms}}

\def\urms{u_{\rm rms}}

\def\half{{\textstyle{1\over2}}}

\def\onethird{{\textstyle{1\over3}}}

\def\quarter{{\textstyle{1\over4}}}
\newcommand{\yapj}[3]{ #1, {ApJ,} {#2}, #3}

\newcommand{\yija}[3]{ #1, {Int.\ J.\ Astrobiol.,} {#2}, #3}

\newcommand{\yana}[3]{ #1, {A\&A,} {#2}, #3}

\newcommand{\ypasj}[3]{ #1, {Publ.\ Astron.\ Soc.\ Japan,} {#2}, #3}

\newcommand{\yaraa}[3]{ #1, {ARA\&A,} {#2}, #3}

\newcommand{\yphl}[3]{ #1, {Phys.\ Lett.,} {#2}, #3}

\newcommand{\ymn}[3]{ #1, {MNRAS,} {#2}, #3}
\newcommand{\ynat}[3]{ #1, {Nature,} {#2}, #3}
\newcommand{\yptrsa}[3]{ #1, {Phil. Trans. Roy. Soc. London A,} {#2}, #3}

\newcommand{\ypre}[3]{ #1, {Phys.\ Rev.\ E,} {#2}, #3}

\newcommand{\yjour}[4]{ #1, {#2}, {#3}, #4}

\title[Magnetic fields with Euler potentials]
{Magnetic field evolution in simulations with Euler potentials}
\author[
]{
Axel Brandenburg$^{1,2}$\\
$^1$NORDITA, AlbaNova University Center, Roslagstullsbacken 23,
SE 10691 Stockholm, Sweden\\
$^2$Department of Astronomy, AlbaNova University Center,
Stockholm University, SE 10691 Stockholm, Sweden}

\date{$Revision: 1.36 $}
\begin{document}

\maketitle

\begin{abstract}
Using two- and three-dimensional hydromagnetic simulations for a range of
different flows, including laminar and turbulent ones, it is shown that
solutions expressing the field in terms of Euler potentials (EP) are in
general incorrect if the EP are evolved with an artificial diffusion term.
In three dimensions, standard methods using the magnetic vector potential
are found to permit dynamo action when the EP give decaying solutions.
With an imposed field, the EP method yields excessive power at small scales.
This effect is more exaggerated in the dynamic case, suggesting an
unrealistically reduced feedback from the Lorentz force.
The EP approach agrees with standard methods only at early times when
magnetic diffusivity did not have time to act.
It is demonstrated that the usage of EP with even a small
artificial magnetic diffusivity does not converge to a proper solution
of hydromagnetic turbulence.
The source of this disagreement is not connected with magnetic helicity
or the three-dimensionality of the magnetic field, but is simply due to
the fact that the nonlinear representation of the magnetic field in terms
of EP that depend on the same coordinates is incompatible
with the linear diffusion operator in the induction equation.
\end{abstract}
\label{firstpage}

\section{Introduction}

In the past few decades magnetic fields have become an integral part of
many branches of observational and theoretical astrophysics.
This is because in virtually all astrophysical bodies the electrical
conductivity is large enough to support electric currents and hence
magnetic fields.
Furthermore, virtually all astrophysical flows produce dynamo action
allowing part of the kinetic energy to be channelled through the
magnetic energy reservoir before it is being dissipated
\citep[see][for a review]{BS05}.
Simulating such flows on the computer can become a serious challenge,
in particular if one wants to reach large magnetic Reynolds numbers and
if one wants to represent huge density contrasts that are typical for
self-gravitating centrifugally supported structures such as galaxies.
The same challenge is met in cosmological simulations that describe the
formation of galaxy clusters and even the formation of galaxies.
Many such simulations have been performed using smoothed particle
hydrodynamics \citep{Dol}.
Its Lagrangian nature is well suited for handling self-gravity \citep{Mon}.
However, incorporating magnetic fields into such simulations has
proved challenging.
A possible solution to this problem may be the use of Euler potentials
\citep{PB07,RP07,RP08}.

On a number of occasions the use of Euler potentials (EP)
has proved useful in astrophysics and magnetohydrodynamics
\citep{Swe50,Dun58,Ste70,Sak79,YL06}.
In this approach the magnetic field is written as
\EQ
\BB=\nab\alpha\times\nab\beta,
\EN
where $\alpha$ and $\beta$ are the EP.
Until recently, the use of EP has only been modestly popular, because
the nonlinearity of such a representation of $\BB$ can lead to difficulties
in representing arbitrary initial conditions.
Furthermore, as pointed out by \cite{Mof78}, magnetic fields with linked or
knotted $\BB$ lines cannot be represented with single-valued differentiable EP.
Another problem is that one has the evolution equations for $\alpha$
and $\beta$ only in the ideal case, i.e.\ when the resistivity vanishes.
In that case one has to solve just two simple advection equations,
\EQ
\DD\alpha/\DD t=0\quad\mbox{and}\quad\DD\beta/\DD t=0.
\EN
Here, $\DD/\DD t=\partial/\partial t+\UU\cdot\nab$ is the advective
derivative and $\UU$ is the velocity.
In recent years the use of Euler potentials has become increasingly
popular in SPH simulations, because
the evolution equations for $\alpha$ and $\beta$ imply that the values
of $\alpha$ and $\beta$ are simply kept fixed at all times.
Several tests have suggested that the use of EP can be superior to solving for
$\BB$ because of the difficulty in preserving $\nab\cdot\BB=0$ in numerical
simulations \citep{DS08}.
Differences between the two results could therefore readily be explained
in terms of $\nab\cdot\BB$ not being zero in the latter approach.
However, this does not eliminate concerns about the correctness of
solutions obtained with the EP method compared to other methods that
also preserve $\nab\cdot\BB=0$.
One such method is to solve for the magnetic vector potential (A method).
In this paper we compare the two methods for a range of different flows.

\section{Euler potentials in simulations}

\cite{Kot09} discussed the fact that the magnetic helicity
vanishes in the EP representation, and so it is clear that this method
is not well suited for studying helical or $\alpha$ effect dynamos
that tend to produce magnetic fields with finite magnetic helicity.
However, we still do not know what the magnetic field will be in such a
case, and whether the EP method can still be useful for studying other
types of dynamos, or at least other types of turbulent magnetohydrodynamic
(MHD) flows.

The goal of this paper is to compare the evolution of the magnetic field
in simulations using the EP method on the one hand and the magnetic
vector potential method (A method) on the other.
We do this by solving the equations for both methods at the same time.
Of course, in the majority of cases, sharp gradients may develop eventually.
This is when numerical methods for solving the equations of {\it ideal} MHD
break down.
It has then been customary to include artificial diffusion in the
evolution equations for $\alpha$ and $\beta$, i.e.\ one considers
solutions of the equations \citep{RP07,RP08}
\EQ
{\DD\alpha\over\DD t}=\eta\nabla^2\alpha,\quad
{\DD\beta\over\DD t}=\eta\nabla^2\beta,
\label{DDalpbet}
\EN
where $\eta$ is the magnetic diffusivity.
In the two-dimensional case with $\BB=\BB(x,y)$ and $B_z=0$ we can
write $\BB=\nab\times(A_z\zzz)$, where $\zzz$ is the unit vector
in the $z$ direction, and $A_z$ obeys the uncurled induction equation,
which can be written as
\EQ
{\DD A_z\over\DD t}=\eta\nabla^2 A_z.
\EN
To compare with the EP method, we choose $\beta=z$ and write
$\AAA=\alpha\nab\beta=\alpha\zzz$, 
where $\zzz$ is the unit vector in the $z$ direction,
so we have $A_z=\alpha(x,y,t)$,
and thus the evolution equation for $\alpha$ becomes identical
to that for $A_z$, even when $\eta\neq0$.
One can also write $\AAA=-\beta\nab\alpha$, which agrees with the
previous formulation after adding the gradient of $\alpha\beta$,
which does not affect the $\BB$ field.

In order to facilitate direct comparison between the EP and A methods,
we solve numerically \Eq{DDalpbet} together with the equation
for the A method (\App{Amethod}),
\EQ
{\DD\AAA\over\DD t}=-\AAA\cdot(\nab\UU)^T+\eta\nabla^2\AAA,
\label{DDAAA}
\EN
where we have assumed $\eta=\const$.
We emphasize that the velocity $\UU$ enters \Eqs{DDalpbet}{DDAAA}
also through the $\DD/\DD t$ derivative, and that the equations for
both approaches are equivalent in the special case of $\eta=0$.
Indeed, if we insert a symmetrized representation,
\EQ
\AAA=\half(\alpha\nab\beta-\beta\nab\alpha)
\label{AAAalphabeta}
\EN
into \Eq{DDAAA}, we obtain
\EQ
\left({\DD\alpha\over\DD t}-\eta\nabla^2\alpha\right)\nab\beta
-\left({\DD\beta\over\DD t}-\eta\nabla^2\beta\right)\nab\alpha
=\RR+\nab\phi,
\label{EPmethod}
\EN
where $\RR$ stands for a residual term, and $\phi$ is given by
\EQ
\phi=\half(\alpha\dot\beta-\beta\dot\alpha)
+\half\nu(\alpha\nabla^2\beta-\beta\nabla^2\alpha)-\UU\cdot\AAA.
\label{phi}
\EN
The $\RR$ term vanishes for $\eta=0$, but
is finite with magnetic diffusivity and is then given by
\EQ
\RR=\eta(\nab\alpha\cdot\nab)\nab\beta
 -\eta(\nab\beta\cdot\nab)\nab\alpha.
\label{Residual}
\EN
Note that the $\nab\phi$ term can be removed from \Eq{EPmethod}
by a gauge transformation; see \App{EulerPot} for the derivation.
However, the $\RR$ term cannot be removed and,
moreover, it has the same highest order of derivatives as the terms
on the LHS of \Eq{EPmethod}, so $\RR$ is in general not small.
This is exactly the reason
why the introduction of artificial diffusion is in general not permissible.
The hope is, of course, that in the limit $\eta\to0$ the EP and A methods
give still reasonably similar results.
In order to illustrate when this is the case, we consider in the following
different flow fields.

\section{Choice of flow fields}

We first consider the case where $\UU$ is a given function
and turn then to the case where $\UU$ is obtained by solving the
momentum and continuity equations.
In the former case we restrict ourselves to flows of the form
\EQ
\UU=\nab\times\psi\zzz+\phi\zzz,
\label{Ugiven}
\EN
where $\psi=\psi(x,y,t)$ and $\phi=\phi(x,y,t)$ are prescribed functions
that will be defined below.
In the latter case we consider the compressible equations with an
isothermal equation of state, so the density $\rho$ is proportional to
the pressure, which is then given by $p=\rho\cs^2$,
where $\cs=\const$ is the isothermal speed of sound.
The governing equations are then
\EQ
{\DD\ln\rho\over\DD t}=-\nab\cdot\UU,
\label{dlnrhodt}
\EN
\EQ
{\DD\UU\over\DD t}=-\cs^2\nab\ln\rho
+\ff+\FF_{\rm visc},
\label{dUdt}
\EN
where 
$\FF_{\rm visc}=\rho^{-1}\nab\cdot2\rho\nu\SSSS$ is the viscous
force, $\nu$ is the kinematic viscosity,
${\sf S}_{ij}=\half(U_{i,j}+U_{j,i})-\onethird\delta_{ij}\nab\cdot\UU$
is the traceless rate of strain tensor, and $\ff$ is a nonhelical random forcing
function consisting of plane transversal waves with random wavevectors
$\kk$ such that $|\kk|$ lies in a band around a given forcing wavenumber
$k_{\rm f}$ \citep{HBD04}.
The vector $\kk$ changes randomly from one timestep to the next.
The forcing amplitude is chosen such that the Mach number $\Ma=\urms/\cs$
is about 0.1.

The total system of equations consists of \Eqs{DDalpbet}{DDAAA} together
with \Eqs{dlnrhodt}{dUdt}.
In all cases the magnetic field is considered infinitesimally
weak, so that the Lorentz force can be neglected.
These equations were solved using the
{\sc Pencil Code}\footnote{http://pencil-code.googlecode.com/} which
is a high-order finite-difference code (sixth order in space and third
order in time) for solving the compressible MHD equations.
The code came with a routine that solves two passive advection--diffusion
equations that were invoked by compiling with {\tt CHIRAL=chiral}, which
is a routine that was originally designed for another purpose to describe
the spontaneous chiral symmetry breaking in biomolecules \citep{BM04}.
Additional diagnostics for monitoring the magnetic field and the current
density have been added to this module for the purpose of this paper.

Initial conditions are generated by setting first
$\alpha$ and $\beta$, and then calculating $\AAA$ from \Eq{AAAalphabeta}.
We consider cubic domains of size $L^3$ using triply-periodic boundary
conditions in all cases, except the first one which is a two-dimensional
case where we assume perfect conductor boundary conditions.
In either case, the magnetic helicity, $H=\int\AAA\cdot\BB\,\dd V$,
is gauge invariant, i.e.\ the transformation $\AAA\to\AAA'+\nab\Lambda$
does not change the value of $H$.
This is because
\EQ
\int\nab\Lambda\cdot\BB\,\dd V=\oint\Lambda\BB\cdot\dd\SSS
-\int\Lambda\nab\cdot\BB\,\dd V
\EN
vanishes owing to the condition $\nab\cdot\BB=0$, and there is no
surface term for periodic domains or perfectly conducting boundaries.
So, the statement that in the EP approach $\AAA\cdot\BB=0$ is merely
a gauge condition \citep{Ste70} does not change the fact that
we always have $H=0$.
On the other hand, the current helicity, $\int\JJ\cdot\BB\,\dd V$,
can well take values different from zero, as has been utilized in the
calculation of force-free equilibria \citep{Sak79}.
In the A approach $H$ can generally be different from zero.
However, if $H=0$ initially, then $H$ can become different from zero
only through resistivity.
This is a consequence of periodic or perfectly conducting boundaries.

\section{Results}

\subsection{Wind-up by a two-dimensional eddy}
\label{WindUp}

We consider first the wind-up of an initially uniform magnetic field,
$\BB=B_0\xxx$, or $\alpha=B_0 y$ and $\beta=z$ in the EP formulation.
We choose a flow with a single eddy given by \Eq{Ugiven}, i.e.\
\EQ
\psi(r)=(U_0/k)\cos^4kr,\quad\phi(r)=\epsilon U_0\psi(r),
\EN
where $r^2=x^2+y^2$ in a domain $-L/2\leq x,y\leq L/2$ and $k=\pi/L$.
For $\epsilon=0$, this flow was used earlier to compute the magnetic
field evolution in ideal MHD \citep{BZ94}, so we were able to compare
our results with theirs in the ideal case.

\begin{figure}\begin{center}
\includegraphics[width=\columnwidth]{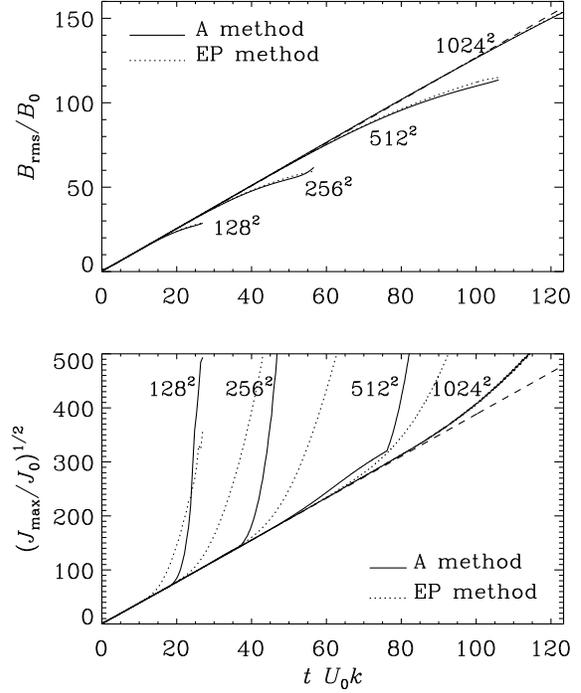}
\end{center}\caption[]{
Evolution of $B_{\rm rms}/B_0$ (upper panel) and
$(J_{\rm max}/J_0)^{1/2}$ (lower panel) for different resolutions
with $\eta=0$ and $\epsilon=2/\pi$.
Here, $J_0=kB_0/\mu_0$ has been used for normalization.
In both plots dashed lines gives the ideal scalings,
i.e.\ linear for $\BB$ and quadratic for $\JJ$.
}\label{pcomp_eddy}\end{figure}

We adopt perfect conductor boundary conditions, which corresponds to
keeping the values of $\alpha$ and $\beta$ on the boundaries equal to
their initial values.
However, it is advantageous to subtract out the linear gradients of
$\alpha=\alpha_0+\alpha_1$ and $\beta=\beta_0+\beta_1$ and solve only
for the departures $\alpha_1$ and $\beta_1$, whose values vanish on
the boundaries.
In our case the imposed gradient fields are $\alpha_0=B_0y$ and $\beta_0=z$,
so the relevant evolution equations are
\EQ
{\DD\alpha_1\over\DD t}=-U_yB_0+\eta\nabla^2\alpha_1,\quad
{\DD\beta_1\over\DD t}=-U_z+\eta\nabla^2\beta_1.
\label{DDalpbet1}
\EN
The result is shown in \Fig{pcomp_eddy} for the ideal case, $\eta=0$,
with $\epsilon=2/\pi$ and different resolution.
One sees clearly that $B_{\rm rms}$ increases linearly with time
while the current density $\JJ=\nab\times\BB/\mu_0$ (with $\mu_0$ being
the vacuum permeability) increases quadratically with time.
In $B_{\rm rms}$ the differences between EP and A methods are small, which is
why we plot in the second panel the maximum value of $|\JJ|$.
Departures from the more accurate solutions obtained at the next higher
resolution appears roughly at the same times, but are seen more clearly
in $J_{\rm max}$ than in $B_{\rm rms}$.
The linear and quadratic scalings for $\BB$ and $\JJ$, respectively,
are well reproduced by either method provided the
resolution suffices to resolve the progressively finer structures as
time goes on.
Looking at the plot of $J_{\rm max}$, one can conclude that
one may need slightly more points with the EP method than with the A method.

\begin{figure}\begin{center}
\includegraphics[width=\columnwidth]{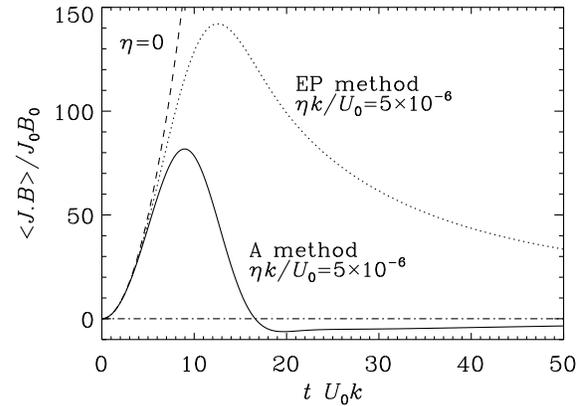}
\end{center}\caption[]{
Evolution of the current helicity $\bra{\JJ\cdot\BB}$ for $\epsilon=2/\pi$
with $\eta=0$ (where A and EP methods both give the same quadratic scaling;
dashed line) and $\eta\neq0$ (where the two methods disagree after some time).
Again, $J_0=kB_0/\mu_0$ has been used for normalization.
}\label{pcomp_JBm_eddy}\end{figure}

For $\epsilon=0$ the EP method gives correct results even in the case
of finite magnetic diffusion, as expected based on the equivalence of
the underlying equations in that case.
This is connected with the fact that the flow is two-dimensional and
confined to the plane only,
However, when $\epsilon\neq0$ we have $U_z(x,y)\neq0$ and $B_z(x,y)\neq0$,
and hence $\beta_1\neq0$.
In this case, the $\RR$ term is in general non-vanishing, and so
\Eqs{DDalpbet}{DDalpbet1} are then no longer equivalent to \Eq{DDAAA},
even though the flow and the field depend only on two spatial coordinates.
This is demonstrated in \Fig{pcomp_JBm_eddy}, where we plot the time
dependence of the current helicity, $\bra{\JJ\cdot\BB}$, in runs with
zero and finite values of $\eta$.
Note the mutual departure of the two methods after some time when $\eta\neq0$.

These experiments have demonstrated that our implementation of the EP
method along with the corresponding diagnostic tools give agreement with
the A method, even when $\epsilon\neq0$, provided $\eta=0$.
However, for $\eta\neq0$ the two methods only agree when $\epsilon=0$.
It may appear that the disagreement is connected with the occurrence of
current helicity, but this is not the case, as will be discussed at the
end of the paper.

\subsection{Roberts flow dynamo}
\label{RobertsFlow}

Next we discuss the \cite{Rob72} flow given by \Eq{Ugiven} with
\EQ
\psi(x,y)=(U_0/k)\cos kx\cos ky,\quad\phi(x,y)=\kf\psi(x,y),
\EN
in the domain $-L/2\leq x,y\leq L/2$, with $k=2\pi/L$ and $\kf=\sqrt{2}k$.
The Roberts flow is one of the simplest flows that produce dynamo action.
The dynamo is however a slow one, i.e.\ its growth rate goes to zero in the
limit $\eta\to0$.
The critical value of $\eta$ is $\eta_{\rm crit}=0.181 U_0/k$, so the
critical value of the magnetic Reynolds number is
$\Rm=U_0 k/\eta_{\rm crit}=5.52$.

\begin{figure}\begin{center}
\includegraphics[width=\columnwidth]{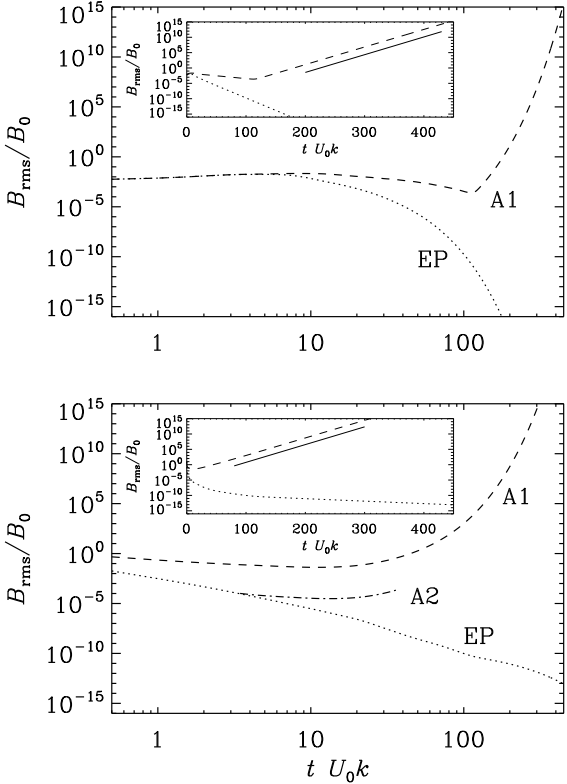}
\end{center}\caption[]{
Comparison of the evolution of $B_{\rm rms}$ in a Roberts flow
for methods~A and EP for a smooth initial condition (upper panel)
and a random one (lower panel) for resolution of $128^3$ meshpoints
and $\Rm=U_0/\eta k=200$.
Both plots are double-logarithmic, so as to see more clearly the mutual
departures of the two solutions at early times.
The insets give the more usual linear--logarithmic representation
showing clearly the exponential growth of the A solution at later times.
The dash-dotted line departing from the EP line at $t=3/U_0k$
is the result of the A method, but with an initial condition calculated
from the EP solution at that time (Run~A2), as opposed to the initial
time (Run~A1).
}\label{pcomp_roberts}\end{figure}

We have considered two different initial conditions, a smooth one given
by $\alpha=\cos ky$ and $\beta=\cos kz$, and a random one where
$\alpha$ and $\beta$ are given by independent random functions.
The results are shown in \Fig{pcomp_roberts}.
For smooth initial fields the EP and A methods agree up to
8 time units ($tU_0k=8$).
This suggests that the EP method gives valid results only when magnetic
diffusion did not yet have time to act.
The A method shows that dynamo action commences after 100 time units,
while the EP method gives only decaying solutions.

For random initial fields dynamo action occurs earlier, after about 10 time
units, but the growth rate is the same as for smooth initial conditions.
The reason for the difference in the onset of the exponential growth is
that the eigenfunction of the dynamo mode overlaps poorly with the smooth
initial condition, and does better so with random initial conditions.
However, for random initial fields (with a spatially white noise power spectrum)
the EP and A methods give different results from the very beginning
(Runs A1 and EP).
This is because of large discretization errors associated with the
numerically different representations of white noise spectra.
In order to check this we have calculated a new initial condition
from the EP solution at time $t=3/U_0k$, when the field has become
sufficiently smooth to be accurately represented by both methods.
Now there is initial agreement, but it is still followed by a departure
immediately afterwards (Run A2).

The results demonstrate quite clearly the difference with the EP method in
handling helical dynamos, just as anticipated previously by \cite{Kot09}.
However, it is still unclear whether the helical Roberts dynamo is just an
exception, or whether the differences are of more general nature.

\subsection{Flows with point-wise zero helicity}
\label{PointWise}

The problem with the Roberts flow is two-fold.
Firstly, it is clear that the dynamo produces a large-scale field
of Beltrami type and is therefore helical.
This is impossible to represent in terms of EP.
Secondly, the dynamo does not exist in the limit $\eta\to0$,
which is the only case where there is hope that the EP method can work.
The latter problem could potentially be alleviated by choosing a flow that
permits fast dynamos, where the growth rate remains finite in the limit
$\eta\to0$.
However, this may not be true if $\eta\to0$ is a singular limit,
which is different from the case $\eta=0$.
Time-dependent flows of \cite{GP92} type tend to be fast dynamos.
An example of such a flow that has also point-wise zero
kinetic helicity is given by \citep[see][]{HCK96}
\EQ
\psi(x,y,t)=\sqrt{3/2}(U_0/k)[\cos kX(x,t)+\sin kY(y,t)],
\EN
\EQ
\phi(x,y,t)=k\sin kX(x,t)\cos kY(y,t),
\EN
\EQ
kX=kx+\cos\omega t,\quad kY=ky+\sin\omega t,
\EN
in the domain $-L/2\leq x,y\leq L/2$, with $k=2\pi/L$.

\begin{figure}\begin{center}
\includegraphics[width=\columnwidth]{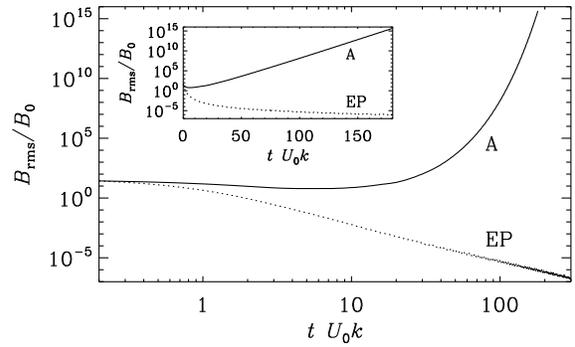}
\end{center}\caption[]{
Comparison of the evolution of $B_{\rm rms}$ for the modified
Galloway--Proctor flow with point-wise zero helicity
for methods~A and EP using $256^3$ meshpoints and $\Rm=10^4$.
Note the power law scaling for the EP method and the exponential
scaling for the A method.
}\label{pt_HCK95_256d}\end{figure}

\begin{figure}\begin{center}
\includegraphics[width=\columnwidth]{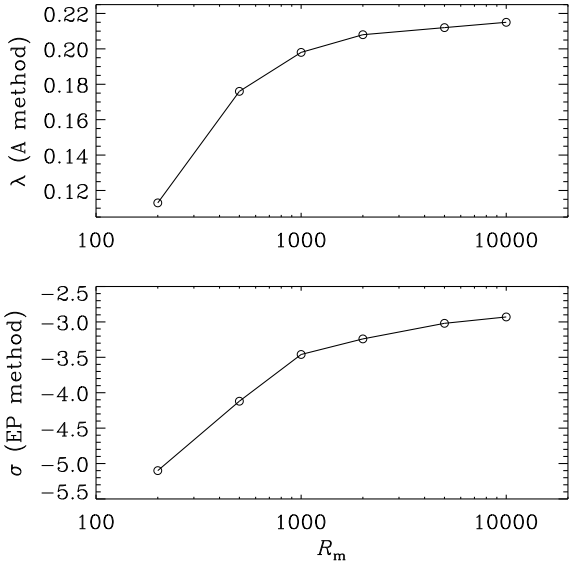}
\end{center}\caption[]{
$\Rm$ dependence of the exponents $\lambda$ and $\sigma$
characterizing the evolution of $\Brms\sim e^{\lambda t}$ for the A method
and $\Brms\sim t^\sigma$ for the EP method for the modified
Galloway--Proctor flow with point-wise zero helicity
for methods~A and EP using $256^3$ meshpoints.
}\label{plamsig}\end{figure}

In \Fig{pt_HCK95_256d} we show an example for $\Rm=U_0/\eta k=10^4$.
Again, it turns out that the EP method does not give solutions that are
compatible with those of the A method.
It turns that, while for the A method the field grows exponentially
like $\Brms\sim e^{\lambda t}$, for the EP method the field decays algebraically
like $\Brms\sim t^{-\sigma}$.
In \Fig{plamsig} we plot the dependence of $\lambda$ and $\sigma$ on $\Rm$.
It turns out that $\lambda$ seems to converge to a finite value
(for the A method), and so does $\sigma$ (for the EP method), confirming
that the functional forms of the time dependencies for the A and EP methods
are indeed different even for large values of $\Rm$.

\subsection{Nonhelically forced isotropic turbulence}
\label{Turbulence}

The flows considered in \Secs{RobertsFlow}{PointWise} are laminar.
Another example of fast dynamo action, where the growth rate is comparable
to the inverse turnover time, is isotropic turbulence.
This is also the example that is closest to the application
to turbulence in galaxy clusters.
In that case there might be a chance to see a tendency toward dynamo
action with the EP method when $\eta\to0$.
Unlike dynamos with helicity, we can only expect the magnetic field to
have length scales smaller than the energy-carrying scale.

\begin{figure}\begin{center}
\includegraphics[width=\columnwidth]{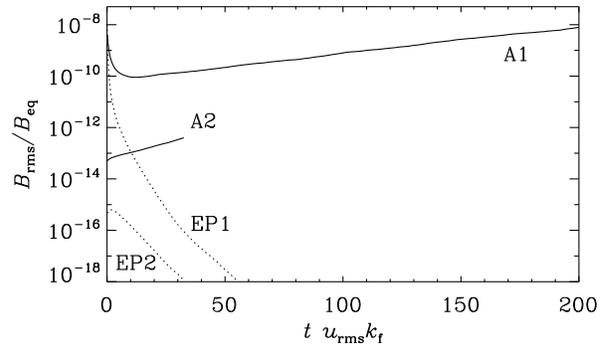}
\end{center}\caption[]{
Comparison of the evolution of $B_{\rm rms}$ in nonhelical turbulence
for methods~A and EP using $128^3$ meshpoints at $\Rm=80$
(Runs A1 and EP1) as well as  $\Rm=160$ (Runs A2 and EP2).
Note that the growth rate for Run A2 is slightly larger than that for A1,
while the decay rates for EP1 and EP2 are the same.
}\label{pcomp_nohel}\end{figure}

\begin{figure}\begin{center}
\includegraphics[width=\columnwidth]{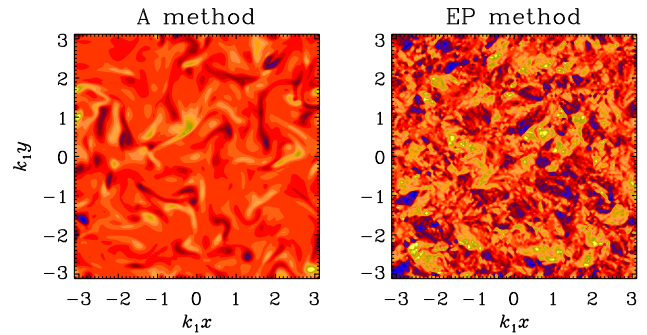}
\end{center}\caption[]{
Comparison of cross-sections of $B_z(x,y)$ for methods~A and EP
from Runs A1 and EP1 after 200 time units,
using $128^3$ meshpoints at $\Rm=80$.
Light (yellow) shades indicate positive values and
dark (blue) shades indicate negative values.
Note the absence of any resemblance between the two fields.
}\label{pslice_128a5}\end{figure}

We consider here the case $\kf/k_1=3$, with $k_1=2\pi/L$
and a magnetic Reynolds number $\Rm=\urms/\eta\kf\approx80$,
using $\nu=\eta$.
The result is shown in \Fig{pcomp_nohel}.
Just like in all previous cases, there is a stark difference in the
evolution of the magnetic field computed with the A and EP methods.
With the A method we reproduce exponential growth consistent with
earlier findings in the literature \citep{CV00,Scheko,HBD03}, while the
EP method gives results that bear no resemblance with those where
small-scale dynamo action is possible.
The same is true of cross-sections of the field; see \Fig{pslice_128a5}.
This strongly suggests that the EP method does not provide a solution
that is close to the expected one, except for the case of a planar flow
that depends only on two coordinates.

\subsection{Spurious growth}
\label{SpuriousGrowth}

\begin{figure}\begin{center}
\includegraphics[width=\columnwidth]{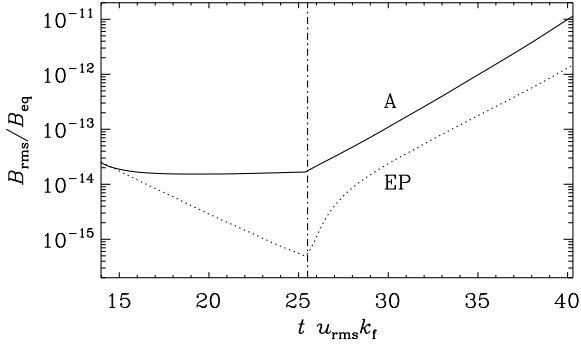}
\end{center}\caption[]{
Spurious growth of $B_{\rm rms}$ to the right of the vertical line
($t\urms\kf>25.5$), for nonhelical turbulence and methods~A and EP
using $128^3$ meshpoints with no resistivity ($\eta=0$), and a
fluid Reynolds number of 80.
}\label{pcomp_nohel2}\end{figure}

In the early days of dynamo theory there have been cases of growing
solutions that later turned out to be spurious due to lack of resolution.
To demonstrate this in the present case, we present in \Fig{pcomp_nohel2}
a solution with $\eta=0$, keeping the fluid Reynolds number equal to 80,
as in \Fig{pcomp_nohel}.

The A and EP solutions show obvious signs of insufficient resolution
with oscillation on the scale of the mesh.
Nevertheless, both solutions show exponential growth with the same growth
rate, which is spurious given the presence of oscillation on the mesh scale.
This illustrates the importance of considering the dependence of
the solutions on $\eta$, as was done in \Sec{Turbulence}.
In that case it turned out that the growth rate increases slightly with
$\Rm$, but this behavior was not reproduced by the EP method.

\subsection{MHD turbulence with imposed field}

The problems considered in \Secss{RobertsFlow}{SpuriousGrowth} had to do with
dynamo action.
This raises the question whether discrepancies between the A and EP methods
also exist in other cases where there is no dynamo action.
As an example we now consider nonhelical turbulence in the presence of an
imposed field using $\alpha_0=B_0y$ and  $\beta_0=z$ as initial fields,
similar what was done in \Sec{WindUp}.
The energy density of the imposed field $B_0$ is comparable to the
kinetic energy density.
This is strong enough to dominate over dynamo action and may even
suppress it.
Here we choose the forcing wavenumber to be $\kf/k_1=1.5$.
The fluid and magnetic Reynolds numbers are again around 80.

\begin{figure}\begin{center}
\includegraphics[width=\columnwidth]{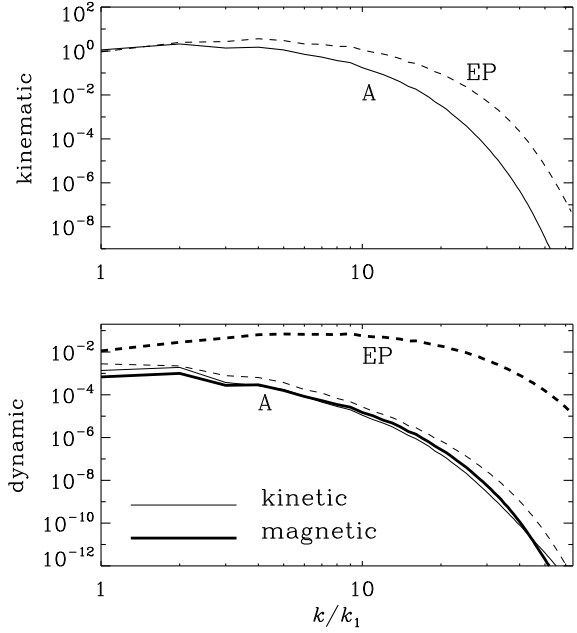}
\end{center}\caption[]{
Magnetic power spectra for the kinematic case (upper panel)
and the dynamic case (lower panel) for the A method (solid lines)
and the EP method (dashed lines).
In the dynamic case the kinetic energy spectra are also shown
as thin solid and dashed lines for the A and EP methods.
Here, $\kf/k_1=1.5$ and $\Rm=80$.
}\label{pcomp_spec}\end{figure}

We consider both the kinematic case without feedback onto the flow and
the dynamic case where the Lorentz force per unit mass, $\JJ\times\BB/\rho$,
has been added to the rhs of \Eq{dUdt} separately for the A and EP methods.
The results are shown in \Fig{pcomp_spec}, where we plot
magnetic power spectra for the two methods.
It turns out that with the EP method, both the kinematic and dynamic
cases yield excessive spectral magnetic energy at smaller scales
(larger wavenumbers) compared to what the A method gives.
Note also that with the EP method the resulting Lorentz force is weaker
than with the A method, making the discrepancy even more pronounced in
the dynamic case.
With the A method kinetic and magnetic energy spectra are in approximate
equipartition with each other, while with the EP method the magnetic field
exceeds the spectral kinetic energy at progressively smaller scale.

\section{Discussion}

Having demonstrated that under a number of circumstances of practical
interest the EP method is unable to provide a meaningful trend when
artificial diffusion is added, one wonders whether the source of this
failure can be identified more precisely.
In particular, we want to know whether this failure is connected with
the inability to represent magnetic fields with helicity, or whether it
is connected with the fact that the magnetic field is three-dimensional.

\begin{figure}\begin{center}
\includegraphics[width=\columnwidth]{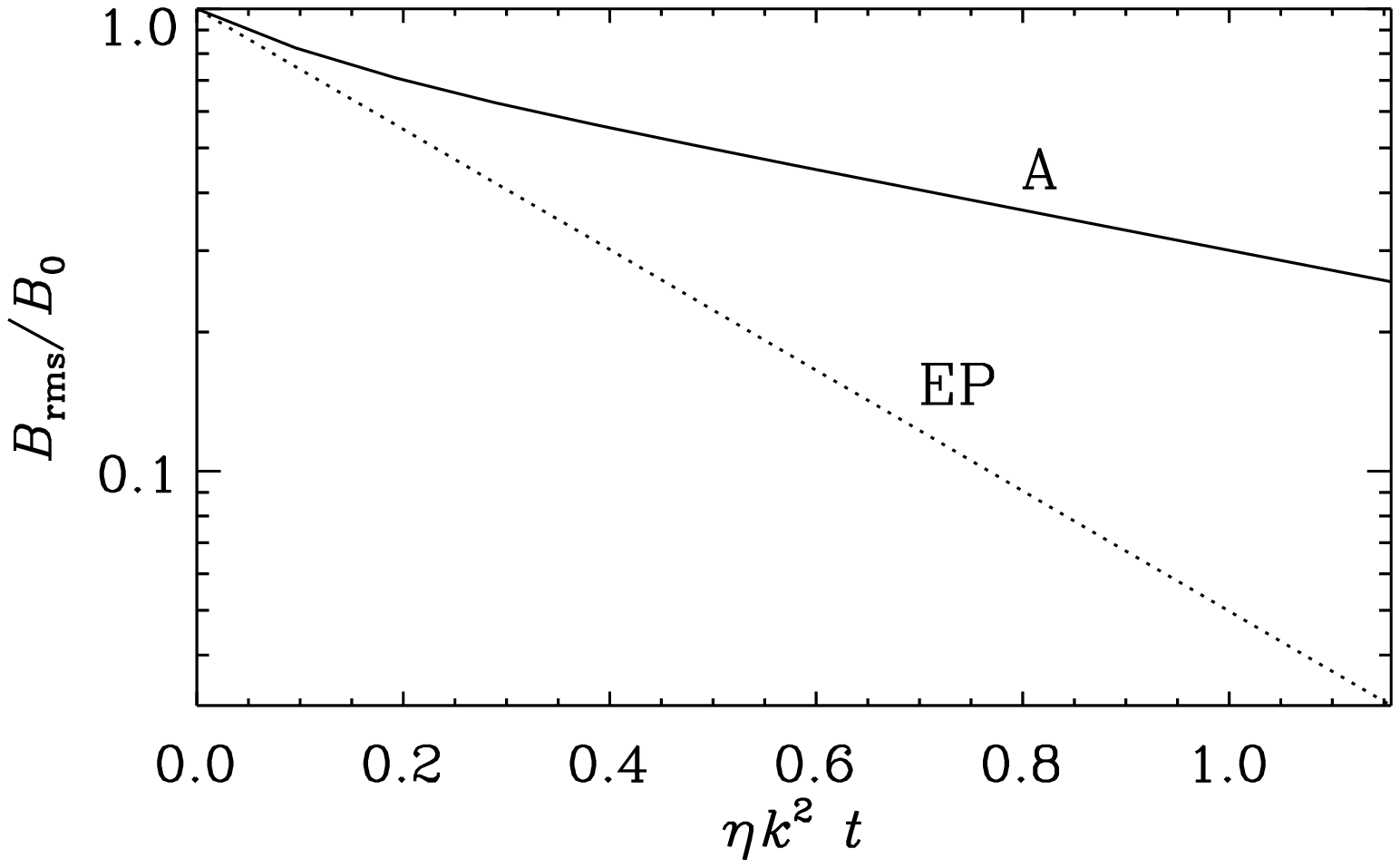}
\end{center}\caption[]{
Decay of $B_{\rm rms}$ for solutions of \Eqs{dAAA}{dalpbet} using as
initial conditions those given by \Eqs{alpbet0}{BB0}, respectively.
}\label{pdecay}\end{figure}

In order to address these issues, we consider now a simple decay problem
with $\UU={\bm0}$ and look for solutions of the equations
\EQ
{\partial\AAA\over\partial t}=\eta\nabla^2\AAA,\quad\mbox{or}\quad
{\partial\BB\over\partial t}=\eta\nabla^2\BB,
\label{dAAA}
\EN
that disagree with solutions of the equations
\EQ
{\partial\alpha\over\partial t}=\eta\nabla^2\alpha,\quad
{\partial\beta\over\partial t}=\eta\nabla^2\beta,
\label{dalpbet}
\EN
even though the initial conditions obey $\BB=\nab\alpha\times\nab\beta$.
The essence of the problem can already be demonstrated with a nonhelical
field in two dimensions.
An example is
\EQ
\alpha=-\cos ky,\quad
\beta=\cos kx\sin ky,
\label{alpbet0}
\EN
which gives
\EQ
\BB(\xx,0)=(0,0,k^2\sin kx\sin^2\!ky)
\label{BB0}
\EN
as initial field.
Note that
\EQ
\mu_0\JJ(\xx,0)=(2\sin kx\cos ky,-\cos kx\sin ky,0)k\sin ky,
\EN
so $\JJ\cdot\BB=0$.
With periodic boundary conditions, \Eq{dalpbet} results in
exponential decay of $\alpha$ and $\beta$ while, owing to the nonlinear
representation of $\BB=\nab\alpha\times\nab\beta$, the $\BB$ field shows
a non-exponential decay; see \Fig{pdecay}.
This problem is also clear from the fact that the $\RR$ term in
\Eq{Residual} does not vanish.
This is generally a consequence of $\alpha$ and $\beta$ being
simultaneously dependent on the same coordinates
(in this case both $\alpha$ and $\beta$ depend on $y$).
Alternatively, if we choose $\alpha=\alpha(y)$ and $\beta=\beta(x)$ with
\EQ
\alpha=\half ky-\quarter\sin2ky,\quad
\beta=\cos kx,
\label{alpbet02}
\EN
which also results in $\BB(\xx,0)$ given by \Eq{BB0}, then $\RR={\bm0}$ and
$\alpha(y,t)$ shows a non-exponential decay---compatible with the correct
solution of $\BB(\xx,t)$.

In general, $\alpha$ and $\beta$ are functions of all three coordinates,
so the $\RR$ term in \Eq{Residual} does not vanish
and the EP method with artificial diffusion will give wrong results.
Thus, we can say that the failure of the EP method in the presence
of artificial diffusion is not related to magnetic helicity nor to the
three-dimensionality of the magnetic field, but simply to the fact
that the nonlinear representation of the magnetic field in terms of
independent functions $\alpha$ and $\beta$ is incompatible with the
linear diffusion operator.

\section{Conclusions}

The EP method can give reliable results when $\eta=0$,
i.e.\ when one is interested in solutions to the ideal MHD equations.
However, in practice this is not possible, especially when the flows
are turbulent, because energy must be dissipated at the smallest scale.
When one allows for magnetic diffusion to be present, there is
no agreement between the EP and A methods.
As a consequence, it is impossible to use the EP method to study dynamos.
Even fast dynamos, which have finite growth rate in the limit $\eta\to0$,
cannot be modelled with the EP method.
This means that any growth of the magnetic field found with the
EP method cannot be due to dynamo action.
This result is not just restricted to helical dynamos that can produce
large-scale fields, but it also applies to nonhelical dynamos that produce
small-scale fields.

Major discrepancies occur even in the case of an imposed field and
in the absence of dynamo action.
It is found that the EP method yields excessive spectral energy,
in particular at small scales.
This discrepancy becomes even more pronounced in the dynamic case
owing to an apparent reduction of feedback from the Lorentz force
compared with the A method.
Indeed, the saturation strength of the magnetic field can be about
20 times larger with the EP method than with the A method.

In the ideal case, the A method may be slightly better suited to deal
with limited numerical resolution than the EP method.
However, once the resolution becomes insufficient, there can be cases
where, in three dimensions, spurious exponential growth occurs with
both methods.
This underlines to necessity of diffusive processes, but with the EP
method this inevitably leads to incorrect solutions.

One might expect that the A method gives good results also in
Lagrangian schemes, because no derivative of $\AAA$ needs to be computed.
An exception is the diffusion term and, of course, the calculation of
$\BB$ and $\JJ$ for the Lorentz force (in full MHD) and for diagnostic
purposes.
The same is true for the EP method as well.
It should therefore be worthwhile to explore the A method also
in Lagrangian schemes.

\section*{Acknowledgments}
I thank Klaus Dolag, Hanna Kotarba, Daniel Price, and Federico Stasyszyn
for discussions about the use of Euler potentials in SPH simulations,
and Matthias Rheinhardt and Karl-Heinz R\"adler for comments on the manuscript.
I acknowledge the use of computing time at the Center for
Parallel Computers at the Royal Institute of Technology in Sweden.
This work was supported in part by
the European Research Council under the AstroDyn Research Project 227952
and the Swedish Research Council grant 621-2007-4064.


\appendix
\section{Derivation of \Eq{DDAAA}}
\label{Amethod}

For completeness we give here the derivation of \Eq{DDAAA}.
The usual equation for $\AAA$ is
\EQ
{\partial\AAA\over\partial t}=-\EE-\nab\phi,
\EN
where $\EE$ is the electric field and $\phi$ is the electrostatic potential.
Using Ohm's law, $\JJ=\sigma(\EE+\UU\times\BB)$, as well as Ampere's law,
$\mu_0\JJ=\nab\times\BB$ and $\BB=\nab\times\AAA$ we have
\EQ
{\partial\AAA\over\partial t}=\UU\times\nab\times\AAA
+\eta(\nabla^2\AAA-\nab\nab\cdot\AAA)-\nab\phi,
\label{dAidt0}
\EN
where we have dropped a term $(\nab\cdot\AAA)\nab\eta$
on the RHS, because in our case $\eta=\const$.
\EEq{dAidt0} can be written as
\EQ
{\partial A_i\over\partial t}=-U_j{\partial A_i\over\partial x_j}
+U_j{\partial A_j\over\partial x_i}
+\eta\nabla^2 A_i-\nab(\eta\nab\cdot\AAA+\phi).
\label{dAidt}
\EN
The first term on the RHS, together with the time derivative
on the LHS, constitute the advective derivative, $\DD\AAA/\DD t$.
Next, we use
\EQ
U_j{\partial A_j\over\partial x_i}=-A_j{\partial U_j\over\partial x_i}
+{\partial U_jA_j\over\partial x_i},
\EN
so we have
\EQ
{\DD\AAA\over\DD t}=-\AAA\cdot(\nab\UU)^T+\eta\nabla^2\AAA
-\nab(\eta\nab\cdot\AAA-\UU\cdot\AAA+\phi).
\EN
After a gauge transformation, $\AAA\to\AAA'+\nab\Lambda$ with
\EQ
\Lambda=\int_0^t\left(\eta\nab\cdot\AAA-\UU\cdot\AAA+\phi\right)\,\dd t'
\EN
we arrive at \Eq{DDAAA}.

\section{Derivation of Equation~(8)}
\label{EulerPot}

In order to verify the $\RR$ term in \Eq{Residual} we calculate
$\nabla^2\AAA$ in terms of $\alpha$ and $\beta$, using
$\AAA=\half(\alpha\nab\beta-\beta\nab\alpha)$, so
\EQA
A_{i,jj}\!&=&\!
\half(\alpha_{,jj}\beta_{,i}-\beta_{,jj}\alpha_{,i})
+(\alpha_{,j}\beta_{,ij}-\beta_{,j}\alpha_{,ij})
\nonumber \\
\!&+&\!
\half(\alpha\beta_{,ijj}-\beta\alpha_{,ijj}).
\label{Aijj}
\ENA
Here, the last term in brackets can be written as the divergence of
$\phi_1=\half(\alpha\beta_{,jj}-\beta\alpha_{,jj})$
minus $\half(\alpha_{,i}\beta_{,jj}-\beta_{,i}\alpha_{,jj})$ which,
in turn, is equal to the first term in \Eq{Aijj}, so we have
\EQ
A_{i,jj}=
(\alpha_{,jj}\beta_{,i}-\beta_{,jj}\alpha_{,i})
+(\alpha_{,j}\beta_{,ij}-\beta_{,j}\alpha_{,ij})+\nabla_i\phi_1.
\EN
The first term in brackets corresponds to the diffusion terms in
\Eq{DDalpbet}, the second term explains the $\RR$ term in \Eq{Residual},
and $\phi_1$ gives one of several terms entering in \Eq{phi}.

For completeness let us here also give the derivation of the remaining
terms.
The time derivative of $\AAA$ is given by
\EQA
{\partial A_i\over\partial t}\!&=&\!\half(
 \dot\alpha\beta_{,i}
+\alpha\dot\beta_{,i}
-\dot\beta\alpha_{,i}
-\beta\dot\alpha_{,i})
\nonumber \\
\!&=&\!\dot\alpha\beta_{,i}-\dot\beta\alpha_{,i}
+\half\nabla_i(\alpha\dot\beta-\beta\dot\alpha),
\label{dAdt2}
\ENA
so we have
\EQ
{\partial\AAA\over\partial t}=\dot\alpha\nab\beta-\dot\beta\nab\alpha
+\nab\phi_2,
\label{dAdt}
\EN
where $\phi_2=\half(\alpha\dot\beta-\beta\dot\alpha)$, and dots denote
partial time derivatives.
Finally, from $\UU\cdot\nab\AAA+\AAA\cdot(\nab\UU)^T$,
we have in components form
\EQ
U_j A_{i,j}+A_j U_{j,i}=U_j(A_{i,j}-A_{j,i})-\nabla_i\phi_3,
\label{AijAji}
\EN
where
\EQ
A_{i,j}=\half(\alpha\beta_{,ij}-\beta\alpha_{,ij})
+\half(\alpha_{,j}\beta_{,i}-\beta_{,j}\alpha_{,i}),
\label{Aij}
\EN
and $\phi_3=\UU\cdot\AAA$.
The first term in brackets of \Eq{Aij} is symmetric in $i$ and $j$,
while the second one is antisymmetric, so only the second one
contributes to $A_{i,j}-A_{j,i}$, giving
$$
\UU\cdot\nab\AAA+\AAA\cdot(\nab\UU)^T
=(\UU\cdot\nab\alpha)\nab\beta-(\UU\cdot\nab\beta)\nab\alpha-\nab\phi_3.
$$
The first two terms explain the advection operator in
\Eq{DDalpbet}, while the last term contributes to
$\phi=\phi_1+\phi_2+\phi_3$ in \Eq{phi}.

\end{document}